\documentclass{aa}
\usepackage{psfig}
\usepackage{epsfig}
\def \sax {BeppoSAX}

\begin{document}

\title{Temperature and total mass profiles of the A3571 cluster of galaxies}

\author{J. Nevalainen\inst{1} 
   \and J. Kaastra\inst{2}
   \and A.N. Parmar\inst{1}
   \and M. Markevitch\inst{3}
   \and T. Oosterbroek\inst{1}
   \and S. Colafrancesco\inst{4}
   \and P. Mazzotta\inst{3}}

\offprints{J. Nevalainen (jnevalai@astro.estec.esa.nl)}

\institute{
       Astrophysics Division, Space Science Department of ESA, ESTEC,
       Postbus 299, NL-2200 AG Noordwijk, The Netherlands
\and   SRON Laboratory for Space Research, Sorbonnelaan 2, 
       3584 CA Utrecht, The Netherlands
\and   Harvard-Smithsonian Center for Astrophysics, 60 
       Garden Street Cambridge, MA 02138, USA
\and   Osservatorio Astronomico di Roma, via dell'Osservatorio 2, I-00040 Monteporzio, Italy}

\date{Received ; Accepted: }

\abstract{We present BeppoSAX results of a spatially resolved spectral analysis of A3571, a relaxed nearby cluster of galaxies. In the central 2\arcmin\ 
(130 ${\rm h_{50}^{-1}}$~kpc) radius the 
metal abundance is $0.49 \pm 0.08$ solar and the absorption $(1.13 \pm 0.28) \times 10^{21}$~atom~cm$^{-2}$, whereas elsewhere within an 8\arcmin\ 
(520 ${\rm h_{50}^{-1}}$~kpc) radius 
the abundance is $0.32 \pm 0.05$ solar and the absorption consistent with the galactic value of $4.4 \times 10^{20}$~atom~cm$^{-2}$. The significant central 
metal abundance enhancement is consistent with the supernova enrichment scenario. The excess absorption may be attributed to the cooling flow, 
whose mass flow rate is $80 \pm 40 \ {\rm M_{\odot}}$~yr$^{-1}$ from our spectral fit. The BeppoSAX and ASCA radial temperature profiles agree over
the entire overlapping radial range r $<$ 25\arcmin\ = 1.6 ${\rm h_{50}^{-1}}$~Mpc.
The combined BeppoSAX and ASCA temperature profile exhibits a constant value out to a radius of $\sim$10\arcmin\ (650 ${\rm h_{50}^{-1}}$~kpc) and a significant decrease 
(${\rm T \propto r^{-0.55}}$, corresponding to $\gamma=1.28$) at larger radii.  These temperature data are
used to derive the total mass profile. The best fit NFW dark matter density model results in a temperature profile that is not convectively stable, but the model
is acceptable within the uncertainties of the data. The temperature profile is acceptably modeled with a ``core'' model for the dark matter density, consisting of a core radius with a constant slope at larger
radii. With this model the total mass and formal 90\% confidence errors within the virial radius r$_{178}$ (2.5 ${\rm h_{50}^{-1}}$~Mpc) are ${\rm 9.1^{+3.6}_{-1.5} \times 10^{14} \ h_{50}^{-1} \
M_{\odot}}$, by a factor of 1.4 smaller than the isothermal value. The gas mass fraction increases with radius, reaching 
f${\rm _{gas}(r_{178}) = 0.26^{+0.05}_{-0.10} \times h_{50}^{-3/2}}$. Assuming that the measured gas mass fraction is the lower limit to the primordial 
baryonic fraction gives ${\rm \Omega_{m} < 0.4}$ at 90\% confidence. }

\maketitle

\markboth{A3571 temperature and total mass profiles}{A3571 
temperature and total mass profiles}

\section{Introduction}
Abell 3571 (hereafter A3571) is a nearby (z=0.04) bright and hot ($\sim$ 7 keV) cluster of galaxies, situated in the Shapley supercluster. 
A3571 is a rich ($>100 $ galaxies) Bautz-Morgan type I cluster (Abell et al. 1989). It is known to possess at its center a giant galaxy 
MCG05-33-002 with extensive optical halo (Kemp \& Meaburn 1991). The galaxy distribution, as well as the optical halo, are aligned in the direction
of the major axis of the central galaxy. Even though the optical data imply dynamical activity, the X-ray observations have revealed that A3571 is a
relaxed cluster. The ROSAT PSPC (position sensitive proportional counter) image of this cluster shows no substructure and no deviations from the azimuthal symmetry (Nevalainen et al. 2000) while 
the ASCA temperature map shows no azimuthal structure indicative of cluster mergers (Markevitch et al. 1998). A3571 has a considerable cooling flow (Peres et
al. 1998), also indicative of a relaxed cluster. 

In this paper we report on a BeppoSAX observation of A3571. We use the data to perform spatially resolved spectroscopy in order to determine the temperature
profile of A3571. Using the BeppoSAX temperature profile derived here, together with the ROSAT brightness profile (Nevalainen et al. 2000), and ASCA temperature profile of A3571 (Markevitch et al. 1998), we constrain the total mass in A3571, assuming hydrostatic equilibrium.

We use $H_{0} \equiv 50 \ {\rm h_{50}}$ km s$^{-1}$ Mpc$^{-1}$, $\Omega_{m} = 1$, $\Omega_{\Lambda} = 0$, which corresponds to the linear scale of 65 kpc/arcmin at the cluster redshift z = 0.04. We consider uncertainties and significances at 90\% level throughout the paper.
 
\section{Observations}

The cluster of galaxies A3571 was observed by BeppoSAX (Boella et al. 1997a) between 2000 February 2 17:42 and February 6 07:41 UTC.
In this {\it paper} we discuss data from the imaging Medium-Energy Concentrator Spectrometer (MECS; 1.8--10~keV; Boella et al. 1997b), and the imaging 
Low-Energy Concentrator Spectrometer (LECS; 0.1--10~keV; Parmar et al. 1997). The MECS consists of two identical grazing incidence telescopes with imaging 
gas scintillation proportional counters in their focal planes. The LECS uses an identical concentrator system to the MECS, but utilizes an ultra-thin 
entrance window and a driftless configuration to extend the low-energy response to 0.1~keV. The fields of view (FOV) of the LECS and MECS are circular with 
diameters of 37\arcmin\ and 56\arcmin, respectively. The MECS has a 2\arcmin\ thick circular window support structure, or strongback, which is centered 
10\arcmin\ from the FOV center. In addition, four 2\arcmin\ thick radial spokes extend outwards from the circular support. The LECS strongback consists of a
square grid with a separation of 4\arcmin. Standard screening and reduction procedures were applied to produce linearized event files using the SAXDAS 2.0.0
package. The exposure times for the MECS and LECS were 66~ks and 28~ks, respectively.

\section{Spatially resolved spectroscopy}

The energy dependent Point Spread Functions (PSF) of the LECS and MECS instruments complicate the analysis of extended objects such  as clusters of galaxies.
In the case of the MECS the problem is less severe, because the mirror and the detector PSF partially cancel each other. For the LECS, the 90\% encircled energy 
radius increases from 6\arcmin\ to 9\arcmin\ between 8~keV and 0.2~keV. In this work, we accounted for the energy dependent PSF by generating appropriate 
instrument response files to be used when fitting the accumulated spectra. For the response generation, we used the ROSAT brightness profile of A3571 
(Nevalainen et al. 2000) to model the flux distribution. For the MECS we produced corrected auxiliary responses with the SAXDAS Effarea tool, as was used
for the clusters A2319 (Molendi et al. 1999) and A2256 (Molendi et al. 2000). Effarea convolves the brightness model with an analytical model for the MECS 
PSF to calculate the instrumental response (see D'Acri et al. 1998 for details). Effarea corrects for the energy-dependent vignetting. For the LECS we used 
the ray-tracing code LEMAT (Lammers 1997) to produce appropriate responses, taking into account the extraction region size and source position on the 
detector, mirror vignetting and obscuration by the strongback. For any given extraction region and energy LEMAT estimates the contribution of photons from 
regions external to the extraction region, and the number of photons originating from within the extraction region that are detected externally (see Kaastra
et al. (1999) for the application of LEMAT in the case of the A2199 cluster). 

\begin{figure*}
\centerline{\psfig{figure=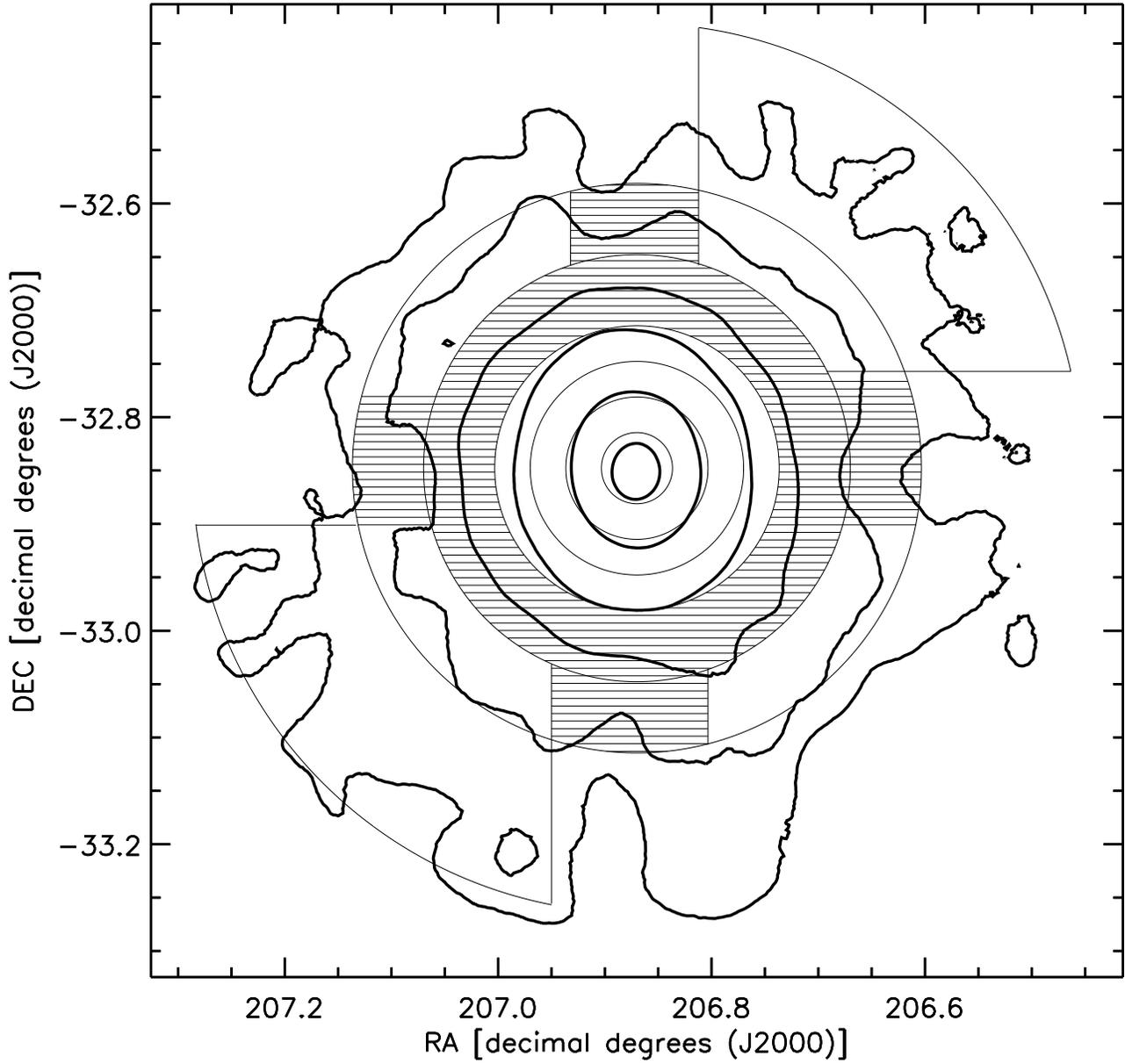,width=18.0cm,angle=0}}
\caption[]{BeppoSAX MECS 1.8--10~keV surface brightness image of A3571. The thick lines are the brightness contours after smoothing with a gaussian of 
$\sigma = 1$\arcmin \ at values of 83, 33, 8, 3, 1 and 0.8\% of the peak value of 1.2$\times 10^{-2}$ c s$^{-1}$ arcmin$^{-2}$. The thin lines show the 
regions where spectra were accumulated. The shaded area was excluded from the accumulations due to obscuration from the strongback}
\label{map_fig}
\end{figure*}

We accumulated LECS and MECS spectra in annular regions centered on the X-ray emission maximum in the MECS (see Fig. \ref{map_fig}). The bounding radii for 
MECS and LECS spectra are 0\arcmin--2\arcmin, 2\arcmin--4\arcmin, 4\arcmin--6\arcmin, 6\arcmin--8\arcmin, and 12\arcmin--16\arcmin. Between 
16\arcmin--25\arcmin\ only MECS data were used due to the smaller LECS FOV. The MECS data between radii of 8\arcmin--12\arcmin \ are shadowed by the strongback,
and the LECS data alone do not give adequate constraints on the spectral parameters. Between radii of 12\arcmin--16\arcmin \ the regions shadowed by the 
strongback are excluded from the MECS spectra. Between 16\arcmin--25\arcmin \ the regions affected by the strongback and the calibration sources are 
excluded. We obtained background spectra from standard blank sky observations using the same extraction regions as for the source spectra. LECS and MECS 
data were selected in the energy ranges 0.1--4.0~keV and 1.8--10~keV, respectively. 

The LECS and MECS spectra were rebinned to oversample the full-width 
half maximum of the energy resolution by a factor 3 and additionally to have a minimum of 20 counts per bin to allow use of the $\chi^2$ statistic, except 
at radii of 12\arcmin--16\arcmin \ and 16\arcmin--25\arcmin \ where a minimum of 30 and 40 counts per bin was used to improve the signal to noise ratio. 
At given radii (below 16\arcmin) the MECS and LECS spectra were fit simultaneously allowing the relative normalization between the two instruments to vary 
by $\pm$30\% to allow for cross-calibration uncertainties (e.g., Fiore et al. 1999). We allowed a 2\% systematic error in the fits to account for absolute
calibration uncertainties.  

\subsection{Radial analysis}
\subsubsection{Global fit}

\begin{figure*}
\centerline{\psfig{figure=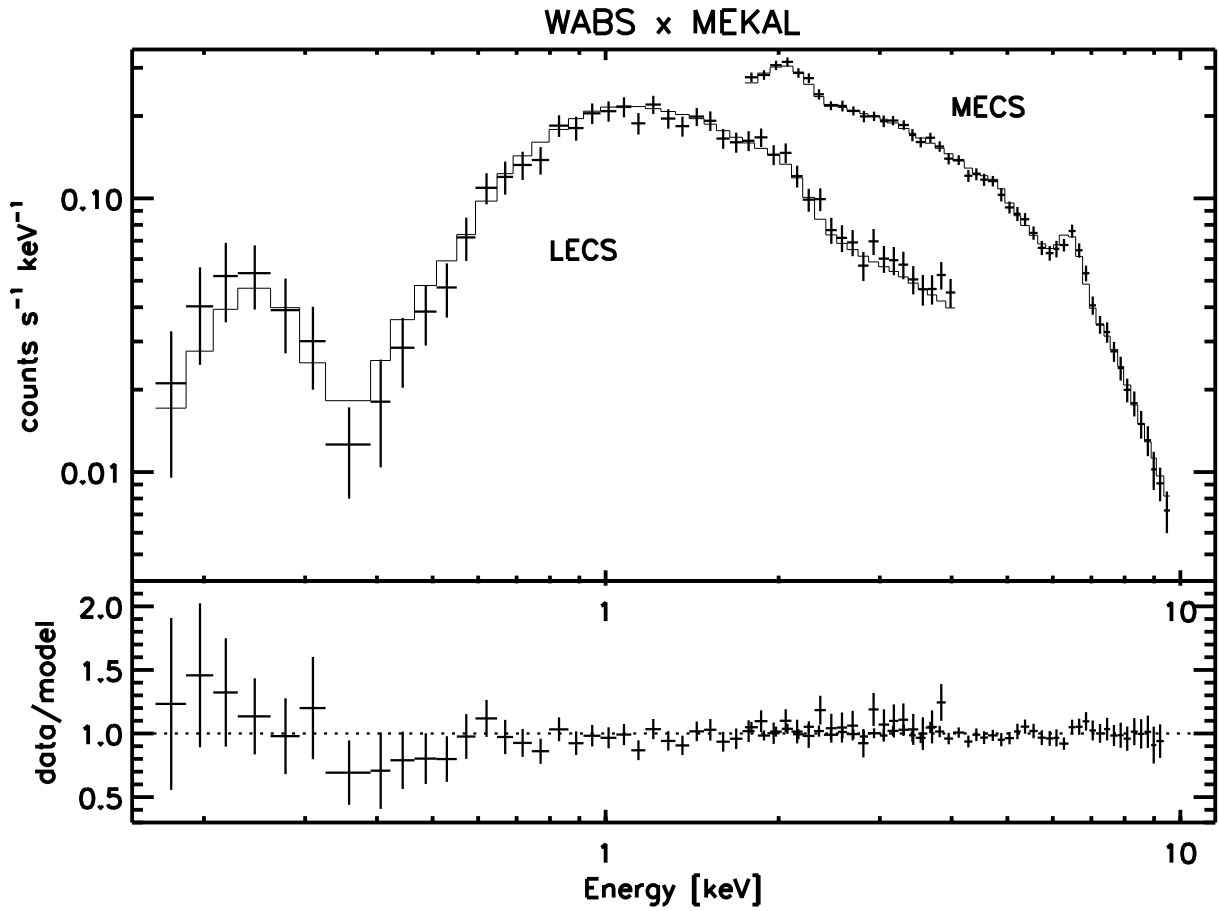,width=18.0cm,angle=0}}
\caption[]{The LECS and MECS spectra accumulated between 2\arcmin--8\arcmin\ together with the best-fit model single component ({\sc mekal}) model absorbed 
by the galactic column. Significant structure is visible at energies $<$0.6~keV}
\label{spectra1_fig}
\end{figure*}

To obtain the global properties of the intracluster gas, we fitted the MECS and LECS data between radii 2\arcmin--8\arcmin. The inner 2\arcmin\ region was 
excluded from the fit, because A3571 is known to have a moderate cooling flow in the center (see below), and the data beyond 8\arcmin \ were excluded due to 
strongback obscuration. We model the emission from the hot intracluster gas using the {\sc mekal} model (Mewe et al. 1995) in XSPEC. 
A single temperature fit to 2--8\arcmin\ spectra with the absorption, ${\rm N_{H}}$ fixed at the galactic value of $4.4 \times 10^{20}$~atom~cm$^{-2}$ 
(Murphy et al. 2000) is unacceptable with a $\chi^2$ of 120.0 for 95 degrees of freedom (dof). There are systematic residuals below 0.3 keV (positive) and between 0.3--0.6~keV (negative), 
(see Fig.~\ref{spectra1_fig}). These features suggest the presence of a warm absorber, which has been detected in several cooling flow clusters 
(Buote 2000a, 2000b). To get an order-of-magnitude estimate of the properties of this component, we added an absorption edge to the model to account 
for the extra absorption between 0.3--0.5~keV and an additional {\sc mekal} 
component to account for the excess emission below 0.3 keV. We kept ${\rm N_{H}}$ fixed at the galactic value and constrained the metal abundances 
of both thermal components to be equal, since the data are not adequate to determine them both separately. The resulting fit is acceptable with a 
$\chi^2$/dof of 90.3/91 and requires an absorption depth of $5.9^{+2.6}_{-2.7}$ at an edge energy of $0.27 \pm 0.01$~keV. 
The warm component has T = $0.18 \pm 0.03$~keV and a normalization 0.27$^{+0.16}_{-0.20}$ that of the hot component. 

However, this model is not physically
meaningful, because the edge energy and the temperature are not linked by physical processes. Therefore we built a warm absorber model by computing the
transmission of the {\sc mekal} model as a function of energy for a grid of temperatures, abundances and column densities, and imported this table model
into XSPEC. We fitted the data with a model consisting of the above warm absorber model, Galactic ${\rm N_{H}}$, a {\sc mekal} component and an 
additional {\sc mekal} component, whose temperature we fixed to that of the absorber. We were not able to explain the observed 
features with this model. To produce the observed depletion at 0.3-0.5 keV the absorber should have a temperature below $\sim$0.01 keV, but at such low
temperatures the model does not produce the observed excess emission at 0.1-0.3 keV. Also, the oxygen absorption edge at 0.6 keV, predicted with this 
model, is not observed. 
The observed features may be due to LECS background or response uncertainties, or due to a more complex absorber. At the lowest energies the background level is comparable to the cluster emission. 
The PSF correction applied to the LECS response is largest at the lowest energies and so the uncertainties may be stronger there. Also, the galactic 
absorption reduces the intrinsic flux by three orders of magnitude at 0.1 keV, and small deviations in the abundances of the absorbing material could 
produce low-energy structure. More complex absorption could result from ionized or partially covered material. However, the quality of the data is 
insufficient to disentangle these effects.

Fitting the data with a model consisting of a Galactic ${\rm N_{H}}$ and a single {\sc mekal} component, excluding bins below 0.6 keV leads to an acceptable
fit (see Table 1). As seen in Fig. \ref{spectra1_fig}, the slight miscalibration of the MECS pulseheight-channel to energy relation results in slight 
offset between the iron 6.7 keV line of the plasma model and the data (as noticed in the case of A2256 by Molendi et al. 2000).
The reason for this is unclear. The in-flight response calibration is made using Crab data, and possible gain problems should have been taken
care of by the response creation procedures. At the moment it is not possible to accurately correct for this effect. However, the mismatch of the iron 
line energy has a negligible effect on temperatures. We verified this by fitting the data excluding the 6 -- 7 keV band and noticing that the changes 
in best-fit temperatures were less than the parameter uncertainties. We estimated the effect of the gain miscalibration on the temperature measurements by
adjusting the gain linear coefficient by 1\%, after which the model iron line matched the data, and redoing the fits with this adjustment (see Table 1).
This does not properly correct for the gain problem, but we use this as an estimate of the systematic uncertainty due to gain 
miscalibration. This adjustement results in systematic shift towards lower temperatures of about the size of the parameter uncertainties at radii below 8\arcmin\ . 
At larger radii the effect is negligible. In the following, when presenting results we include this uncertainty in the parameter values. 

\subsubsection{Center}
In the 0\arcmin--2\arcmin\ radial bin the spectral fit is formally acceptable with a model consisting of a single absorber and a single temperature plasma 
component with Galactic ${\rm N_{H}}$. However, the fit improves significantly, when ${\rm N_{H}}$ is allowed to be a free parameter. 
The resulting absorption ($(11.3^{+2.7}_{-2.6}) \times 10^{20}$~atom~cm$^{-2}$) is significantly (by a factor of 2.6) above the galactic value. Qualitatively similar behavior has been found in other cooling flow clusters (e.g. Allen \& Fabian 1997; Molendi et al. 1998; Molendi \& De Grandi, 1999; De Grandi \& Molendi 1999b). 
The corresponding BeppoSAX metal abundance (0.49$\pm 0.08$ in solar units) is significantly (by a factor of 1.5) above the global value. A study of several 
clusters, including A3571 using ASCA data (Dupke \& White 2000) also shows central metal enhancement and gives for A3571 abundance values consistent with 
ours. 

The cooling flow found in the center of A3571 (Peres et al. 1998) contributes a comparable amount of flux to the ambient gas in the central bin. To study 
this more we added a cooling flow model ({\sc mkcflow} in XSPEC) fixing the lower boundary of the continous temperature range to 
0.1~keV and forcing the upper boundary to equal the ambient temperature. The metal abundance
was constrained to be equal to the ambient one, because the data quality does not allow them to be measured separately. We also fixed the mass flow rate to
75 ${\rm M_{\odot} \ yr^{-1}}$ as found in ROSAT analysis (Peres et al. 1998). As a result, the temperature of the ambient gas in the central bin becomes
$9.5^{+1.1}_{-1.5}$~keV. Letting the mass flow rate to be a free parameter, the temperature of the gas becomes very uncertain, but nevertheless the 
cooling flow is significantly detected, i.e. $\dot{M} = 80 \pm 40 \ {\rm M_{\odot} yr^{-1}}$ at 90\% confidence level.
 
\begin{table*}
\caption[]{Fit results. The model used in the fits is {\sc mekal} absorbed by Galactic ${\rm N_{H}}$, unless stated differently in remarks column.
The errors are given at 90\% confidence level}
\begin{flushleft}
\begin{tabular}{lccccccl}
\hline\noalign{\smallskip}
 & \multicolumn{3}{c}{NO GAIN CORRECTION} & \multicolumn{3}{c}{GAIN CORRECTION} & \\
Radius    &     T     & Abund    & $\chi^2$/dof  &      T     & Abund    & $\chi^2$/dof & Remarks\\
(\arcmin) & (keV)     & (solar)  &               &    (keV)   &          &              &  \\          
\hline\noalign{\smallskip}
2--8      & 7.9$^{+0.4}_{-0.2}$    &   0.32$^{+0.05}_{-0.04}$   &  98.5/82  &   7.6$^{+0.3}_{-0.3}$    &    0.31$^{+0.05}_{-0.03}$     &    95.2/81  &  \\ 
0--2      & 8.1$^{+0.5}_{-0.4}$    &   0.50$^{+0.08}_{-0.07}$   &  87.7/82  &   7.7$^{+0.5}_{-0.3}$    &    0.49$^{+0.09}_{-0.07}$     &    87.7/82  &  wabs $\times$ mekal, ${\rm N_{H}} \equiv$ Galactic\\ 
0--2      & 7.6$^{+0.5}_{-0.5}$    &   0.49$^{+0.08}_{-0.07}$   &  66.8/81  &   7.2$^{+0.4}_{-0.5}$    &    0.49$^{+0.07}_{-0.08}$     &    61.4/81  &  wabs $\times$ mekal, ${\rm N_{H}}$ free \\ 
0--2      & 9.5$^{+1.1}_{-0.9}$    &   0.53$^{+0.09}_{-0.08}$   &  56.1/81  &   8.7$^{+0.9}_{-0.7}$    &    0.52$^{+0.06}_{-0.08}$     &    53.1/81  &  wabs $\times$ (mekal + mkcflow), ${\rm N_{H}}$ free \\ 
2--4      & 7.6$^{+0.5}_{-0.3}$    &   0.31$^{+0.06}_{-0.05}$   &  97.8/83  &   7.3$^{+0.4}_{-0.3}$    &    0.32$^{+0.05}_{-0.06}$     &    92.2/83  &  \\ 
4--6      & 7.8$^{+0.5}_{-0.4}$    &   0.37$^{+0.07}_{-0.07}$   &  95.3/81  &   7.4$^{+0.5}_{-0.4}$    &    0.37$^{+0.07}_{-0.07}$     &    92.9/81  &  \\ 
6--8      & 8.4$^{+0.8}_{-0.7}$    &   0.26$^{+0.09}_{-0.10}$   &  66.4/82  &   8.0$^{+0.8}_{-0.6}$    &    0.25$^{+0.09}_{-0.09}$     &    68.7/82  &  \\ 
12--16    & 6.6$^{+1.1}_{-1.0}$    &   0.32$^{+0.21}_{-0.20}$   &  75.8/71  &   6.4$^{+1.1}_{-1.0}$    &    0.29$^{+0.21}_{-0.19}$     &    76.3/71  &  \\ 
16--25    & 4.4$^{+1.5}_{-1.0}$    &   0.13$^{+0.37}_{-0.13}$   &  61.2/47  &   4.2$^{+1.4}_{-0.9}$    &    0.19$^{+0.41}_{-0.19}$     &    60.8/47  &  \\ 
\noalign{\smallskip\hrule\smallskip}
\end{tabular}
\end{flushleft}
\label{res_tab}
\end{table*}

\subsubsection{Other radii}
\begin{figure*}
\centerline{\psfig{figure=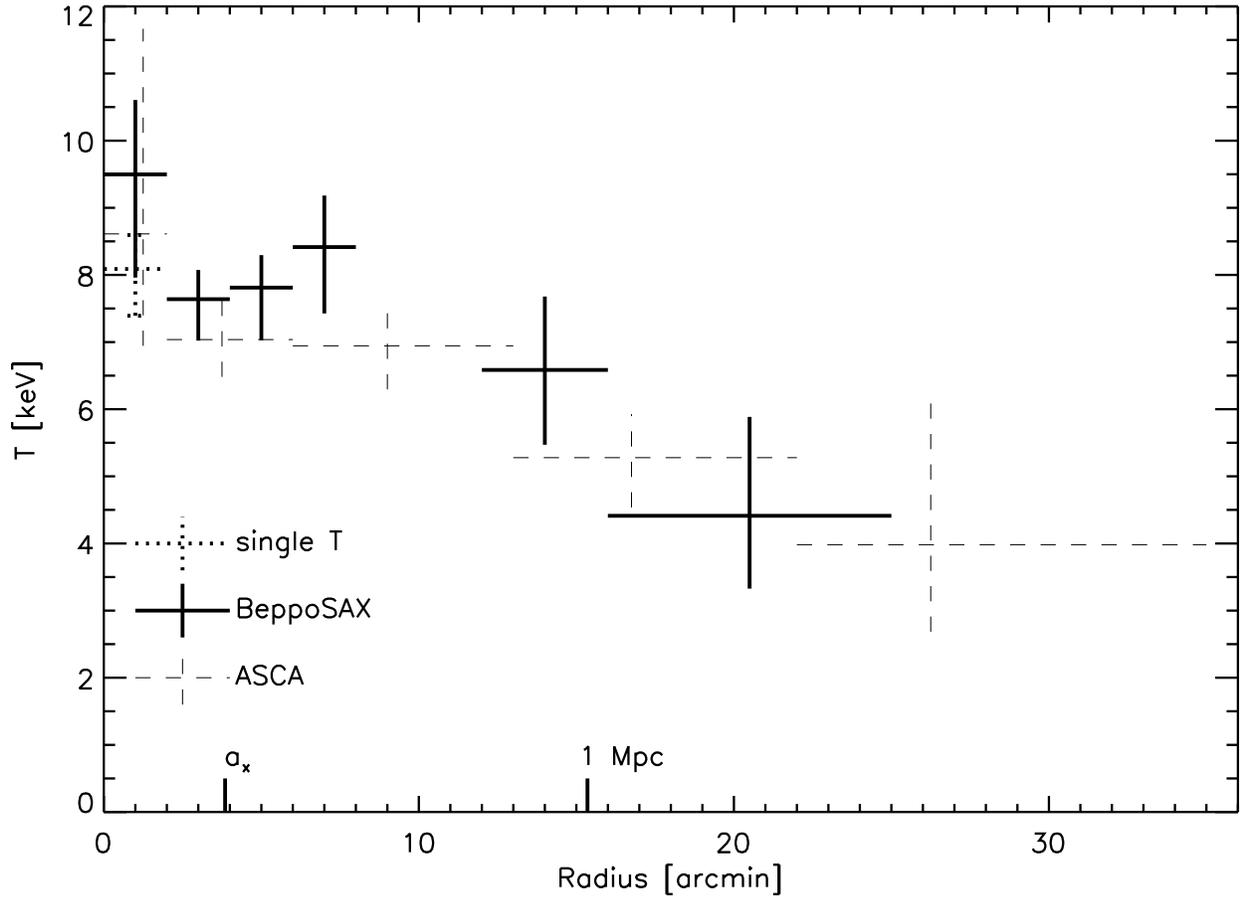,width=18.0cm,angle=0}}
\caption[]{The solid crosses show the best-fit A3571 temperature with 90\% confidence uncertainties obtained from BeppoSAX data using the {\sc mekal} model.
The error bars are expanded to include the shift due to the gain calibration uncertainty (i.e. the uncertainties include the 90\% confidence intervals 
of the values in columns 2 and 5 at Table 1). At radii $>$2\arcmin\ the absorption consists of a galactic 
absorber. In the fit to the central 2\arcmin\ data, ${\rm N_{H}}$ was allowed to vary and a cooling flow component was included. The dotted line shows 
the single temperature fit. The dashed crosses show the values obtained using ASCA by Markevitch et al. (1998)}
\label{Tplot_fig}
\end{figure*}

Using a single {\sc mekal} model (as explained above), we find that temperatures and metal abundances in the radial bins 2\arcmin--4\arcmin, 
4\arcmin--6\arcmin\ and 6\arcmin--8\arcmin\ are consistent with being constant (see Fig.~\ref{Tplot_fig} for the T profile). Between 12\arcmin--16\arcmin\ 
the best fit temperature is lower, but still agrees within the uncertainties with the 2\arcmin--8\arcmin\ fit results. In the 16\arcmin--25\arcmin\ bin the 
temperature is significantly lower than in the other bins. The BeppoSAX temperature profile of A3571 derived here has slightly higher values than
that derived using ASCA for this cluster (Markevitch et al. 1998), but both agree within 90\% confidence uncertainties at each radius.

To quantify the behavior of the temperature profile, we modeled the 3-dimensional temperature profile with a constant value ${\rm T_{const}}$ out to r = 10\arcmin,
and with a behavior ${\rm T \propto r^{-i}}$ at larger radii. We projected this model into BeppoSAX and ASCA bins and iteratively determined the best-fit parameters to the data and to 
the Monte - Carlo sets (see below) in order to obtain the uncertainties. We obtain ${\rm T_{const} = 7.8^{+0.5}_{-0.4}}$ and i$ = 0.55^{+0.27}_{-0.24}$ which 
corresponds to a polytropic model ${\rm T \propto \rho_{gas}^{\gamma - 1}}$ with $\gamma = 1.28^{+0.14}_{-0.12}$ beyond r = 10\arcmin. However, because the 
observed T is constant up to r = 10\arcmin\ and therefore different from the gas density profile, the polytropic model does not give an acceptable fit to the 
data.

Due to the strongback obscuration and low count rates at large radii the BeppoSAX data are not ideal for measuring the spatial variation of the spectral 
parameters in A3571. However, we divided the 2\arcmin--8\arcmin\ MECS and LECS data into four sectors and between radii 12\arcmin\ and 25\arcmin\ we used 
two sectors of MECS data. The temperatures and metal abundances in these sectors are consistent with the average values, i.e. we find no spatial variation. 
This is consistent with the ASCA results of A3571 (Markevitch et al. 1998).  

\subsubsection{Fluxes}
With the best-fit models (using the MECS normalizations) we determined the fluxes and luminosities in complete annuli in different radial bins. Above 
8\arcmin, where the sky coverage is incomplete due to exclusion of strongback and calibration source regions, we assume azimuthal symmetry when computing 
the fluxes. In the 2\arcmin--8\arcmin\ region we measure unabsorbed 2-10 keV fluxes and luminosities of
${\rm F_{2-10} = 7.3 \pm 0.4 \ 10^{-11}}$~erg~s$^{-1}$~cm$^{-2}$ and 
${\rm L_{2-10} = 5.1 \pm 0.2 \ h_{50}^{-2} \ 10^{44}}$~erg~s$^{-1}$ 
and bolometric values of
${\rm F_{bol} = 15.5 \pm 0.7 \ 10^{-11}}$~erg~s$^{-1}$~cm$^{-2}$, 
${\rm L_{bol} = 10.8 \pm 0.4 \ h_{50}^{-2} \ 10^{44}}$~erg~s$^{-1}$. 
For comparison, we distribute the measured ROSAT flux of A3571 (Markevitch 1998) between different radii using the best-fit $\beta$ profile from 
Nevalainen et al. (2000). Between 2\arcmin--8\arcmin\ the 0.1--2.4~keV ROSAT flux is $5.9 \pm 0.3 \ 10^{-11}$~erg~s$^{-1}$~cm$^{-2}$ and agrees with the 
BeppoSAX value of $5.8 \pm 0.1 \ 10^{-11}$~erg~s$^{-1}$~cm$^{-2}$. Also within the 12\arcmin--16\arcmin\ annulus the ROSAT value 
of $0.87^{+0.04}_{-0.05} \ 10^{-11}$~erg~s$^{-1}$~cm$^{-2}$ is consistent with our value of 
$0.84^{+0.07}_{-0.07} \ 10^{-11}$~erg~s$^{-1}$~cm$^{-2}$. 
At the largest radii there is some model dependent discrepancy, because the ROSAT flux estimate is obtained with a single temperature model, whereas the 
BeppoSAX temperature decreases at large radii. However, the ROSAT value for radii of 
16\arcmin--25\arcmin\ ($0.96^{+0.05}_{-0.04} \ 10^{-11}$~erg~s$^{-1}$~cm$^{-2}$) 
agrees with our value
($1.4^{+0.3}_{-0.2} \ 10^{-11}$~erg~s$^{-1}$~cm$^{-2}$) 
within 3$\sigma$ confidence.

\section{Mass determination}
\label{mass}

\subsection{Method}

The ASCA temperature map (Markevitch et al. 1998) and the ROSAT brightness distribution (Nevalainen et al. 2000) of A3571 show no deviations from azimuthal 
symmetry and no substructure, and the BeppoSAX data do not contradict this view. Therefore we are justified in assuming that the hot X-ray emitting material
in A3571 is in hydrostatic equilibrium. 

Since the BeppoSAX temperature profile agrees with the ASCA profile, we use both simultaneously to constrain the total mass. For the details of 
the method we refer to the similar analysis of the ASCA data of A3571 (Nevalainen et al. 2000) and A401 (Nevalainen et al. 1999). Briefly, we model the dark
matter density with two different types of profiles. In the ``core'' model 
\begin{equation}
\ \rho_{{\rm dark}} \propto \left(1 + \frac{r^2}{a_d^2}\right)^{- \alpha/2}
\label{coremodel}
\end{equation}
the density is relatively constant inside the core radius and has a constant slope $\alpha$ at large radii, analoguous to the $\beta$ model for the gas 
density.  
The ``cusp'' profile
\begin{equation}
\rho_{{\rm dark}} \propto \left(\frac{r}{a_{d}}\right)^{- \eta} \left(1 + \frac{r}{a_{d}}\right)^{\eta - \alpha}.
\label{cuspmodel}
\end{equation}
has a changing slope and a central cusp. With $\eta = 1$ and $\alpha = 3$, the cusp model corresponds to the ``universal density profile'' which  is a good description of the cluster cold dark matter halos in simulations of hierarchical clustering
(Navarro et al. 1995, 1997, hereafter NFW). Our data are not adequate to determine the shape of the cusp, and therefore we fix the shape parameter 
$\eta$ to be 1 in the cusp models, as suggested by the above simulations, but vary the other parameters. We solve the hydrostatic 
equilibrium equation (e.g., Sarazin 1988):
\begin{eqnarray}
%\begin{equation}
M_{{\rm tot}}(\le r) = 3.70 \times10^{13} M_{\odot} {T(r) \over {\rm keV}} {r \over {\rm Mpc}} \times  & & \nonumber \\
\left( - {{d \ln{\rho_{{\rm gas}}}} \over {d \ln{r}}} - {{d \ln{T}} \over {d \ln{r}}} \right), & & 
\label{hydreq}
%\end{equation}
\end{eqnarray}
(using $\mu = 0.60$), with respect to $T$, in terms of the dark matter and gas density profile parameters (the analytical solutions are given in Markevitch \& Vikhlinin 1997).
We fix the gas density to that found from the ROSAT analysis (i.e. $\beta$ model with slope parameter $\beta = 0.68\pm 0.03$ and core radius a$_{x} = 3.85\pm0.35$\arcmin\ which describes well the PSPC surface brightness profile data in the radial range considered in this work 
(Nevalainen et al. 2000), calculate the 3-dimensional temperature profile model corresponding 
to given dark matter parameters, project it on the BeppoSAX and ASCA annuli, compare these values to the observed temperatures and iteratively determine the 
dark matter distribution parameters. 

To propagate the errors of the temperature profile data to our mass values, we repeat the procedure for a large number of Monte - Carlo temperature profiles 
with added random errors. The gain calibration uncertainty will only affect the BeppoSAX temperatures and its effect on the combined BeppoSAX + ASCA mass
profile is negligible and we thus ignore it. 
From the distribution of the acceptable Monte - Carlo models, we determine the 1$\sigma$ confidence intervals of the mass 
values as a function of radius. We convert these values to 90\% confidence values, assuming a Gaussian probability distribution. 

\subsection{Results}

Even though our BeppoSAX temperature profile is consistent with the ASCA profile (Markevitch et al. 1998), the BeppoSAX profile is somewhat steeper. As a 
consequence the best-fit cusp model becomes too steep to be convectively stable above r = 26\arcmin , i.e. the effective polytropic index
\begin{equation}
\gamma(r) = {{d \ \log \ T(r)} \over {d \ \log \ \rho(r)}} + 1
\end{equation}
exceeds $\frac{5}{3}$. However, some of the Monte - Carlo models, which quantify the effect of the model parameter uncertainties, are convectively stable. 
Therefore the cusp model, as well as the NFW model are only marginally acceptable, and thus we report no mass values based on this model.

\begin{figure*}
\centerline{\psfig{figure=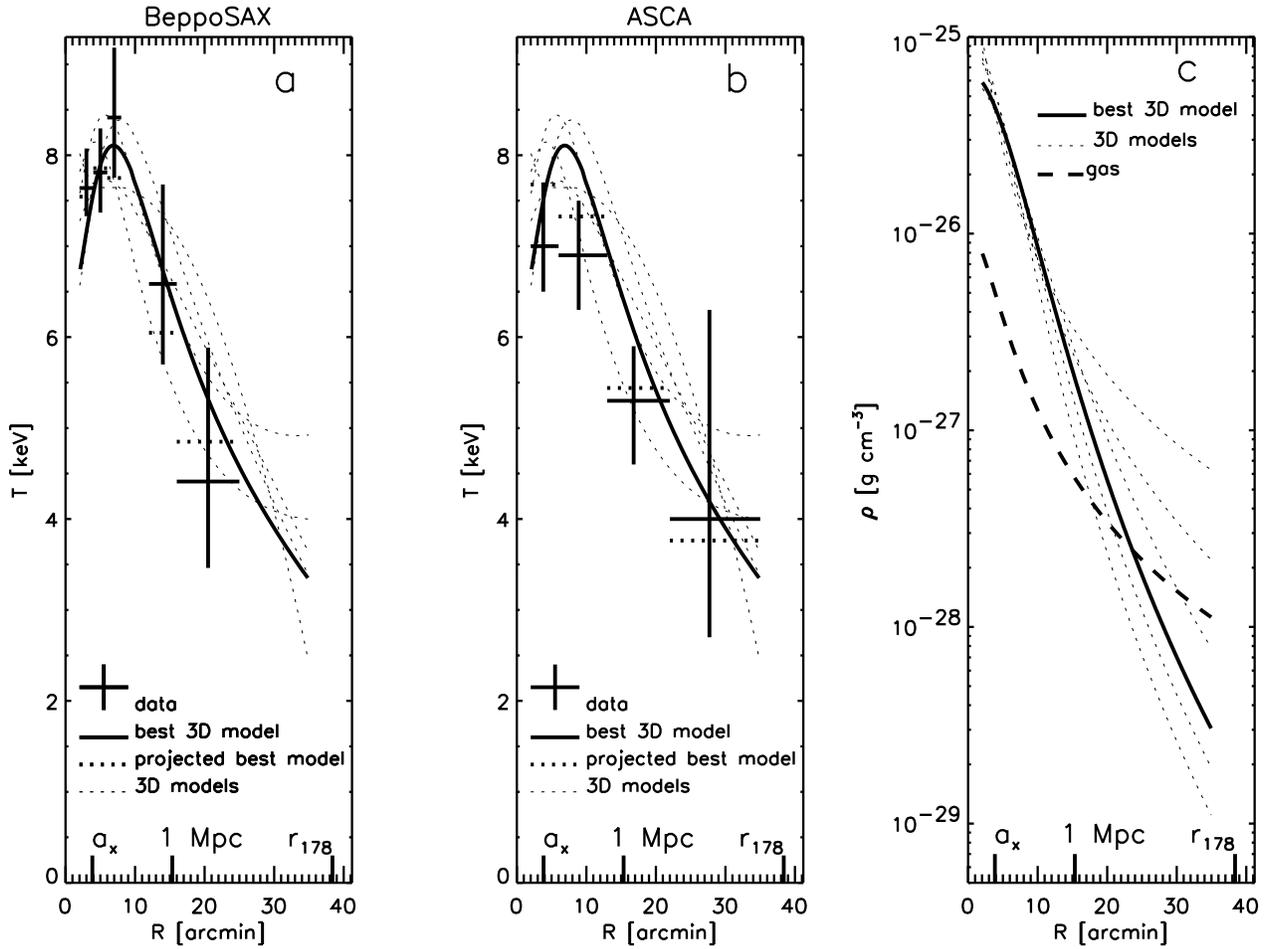,width=18.0cm,angle=0}}
\caption[]{The solid lines in panels a and b show the 3-dimensional best-fit core model temperature. The thick dotted lines in panels a and b show the 
best-fit core model temperatures projected onto the BeppoSAX and ASCA radial bins. The crosses are the temperature data as in Fig.~3. The thin dotted lines 
show a representative sample of Monte - Carlo models. Panel c shows the corresponding dark matter density profiles
(solid and dotted lines) and the gas density profile (dashed line)}
\label{trhoplot_fig}
\end{figure*}

The best-fit core model temperature profile is acceptable with a $\chi^{2}$ of 8.5 for 4 dof. This model is on the border of being convectively unstable, 
but can still be accepted (see Fig. \ref{trhoplot_fig} for the profile). 
The best-fit is obtained with 
a core radius of ${\rm a_{d}}$ = 10\farcm1, a slope of $\alpha = 6.0$, a central temperature of $T_{0} = 6.2$ keV and  central dark matter density
of ${\rm \rho_{d}(0) = 6.6 \times 10^{-26}}$~g~cm$^{-3}$ (6 times as high as the central gas density). The temperature data covers the radial range up to 
35\arcmin. Extrapolating the temperature profile slightly (to $39'$) using our best fit model, we can determine the cluster mass at the
cosmologically interesting radius, $r_{178}$, i.e. the virial radius in ($\Omega_m=1, \Omega_{\Lambda}=0$) universe
based on the spherical collapse model (Eke et al. 1996).
We obtain
\begin{equation}
{\rm r_{178}} = 38.4\arcmin = 2.5 \ h_{50}^{-1} \ {\rm Mpc}
\end{equation}
and the total mass within this radius is
\begin{equation}
{\rm M_{178}} = 9.1^{+3.6}_{-1.5} \times \ 10^{14} \ h_{50}^{-1} \ {\rm M}_{\odot},
\end{equation}
(see Table~\ref{mass_tab} for the mass values at several interesting radii). We note that half of the Monte - Carlo fits used for the mass uncertainty estimation are convectively unstable. 
Models with larger negative temperature gradients at a given radius are more massive if the temperatures at that radius are equal (see Eq. \ref{hydreq}).
Thus, applying the convective stability constraint would produce a strong constraint for the upper bound to the mass profile  
(e.g. at ${\rm r_{178}}$ we would obtain ${\rm M}_{178} = 9.1^{+0.7}_{-1.8} \times \ 10^{14} \ h_{50}^{-1} \ {\rm M}_{\odot}$). However, such a constraint is model dependent and 
does not allow the uncertainties in the temperature measurements to be propagated to the mass estimates. We therefore choose not to use this constraint
to reject any of the Monte - Carlo fits. 

The mass values obtained here are consistent with the corresponding values obtained using only ASCA 
(Nevalainen et al. 2000), and more precise. In addition to the better data,
the improvement in the precision is due to the fact that in the present work we 
cannot apply the cusp model (due to convective instability) as in Nevalainen et al. (2000). However, if we were to apply the convective stability constraint for the Monte - Carlo fits with the core model, the mass uncertainties derived here would be much smaller. 
Our mass values within r$_{500}$ and $r_{178}$ are smaller by factors of 1.1 and 1.4, respectively,  than the mass obtained assuming that the cluster is
isothermal with the emission-weighted ASCA value $T_{X}$ = 6.9 keV (Markevitch 1998) but consistent within the uncertainties. 
Within r$_{500}$ our total mass value is significantly smaller (by a factor of 1.6) than the value predicted by the cosmological simulations of Evrard et al. (1996) for a cluster with T = 6.9 keV.

Our results indicate that the dark matter density falls as r$^{-6}$ whereas the gas density falls as  r$^{-2}$ and thus the gas mass fraction increases rapidly at large radii. At 
r$_{500}$ our gas mass fraction value (see Table~2) is consistent with those for A2256 (Markevitch \& Vikhlinin 1997), A401 (Nevalainen et al. 1999), 
A2199 and A496 (Markevitch et al. 1999) and for samples in Ettori \& Fabian (1999) and Mohr et al. (1999). At the virial radius we obtain
\begin{equation}
f_{gas}(\le r_{178}) = 0.26^{+0.05}_{-0.10} \ h_{50}^{-3/2}.
\end{equation}
Assuming that this is a lower limit on the cosmological baryon fraction, or that 
\begin{equation}
{\Omega_{b} \over \Omega_{m}} \equiv {{\rm M}_{b} \over {\rm M_{tot}}} \ge f_{gas}(\le r_{178}),
\end{equation}
 and using ${\rm \Omega_{b} = 0.076 \pm 0.007}$ (Burles \& Tytler 1998) and ${\rm H_{0} \ge 50}$~km~s$^{-1}$~Mpc$^{-1}$ we obtain an 
upper limit for the cosmological mass density parameter of ${\rm \Omega_{m} \le 0.41}$.

\begin{table}
\caption[]{Mass determinations for A3571 assuming 
${\rm H_{0} \equiv 50 \ h_{50}}$~km~s$^{-1}$~Mpc$^{-1}$.
All uncertainties are formal 90\% confidence limits resulting from our analysis}
\begin{flushleft}
\begin{tabular}{lr}
\hline\noalign{\smallskip}
Parameter & Value \\
\hline\noalign{\smallskip}
M${\rm_{tot}( < 3.85\arcmin = a_{x}}$  ) [${\rm M_{\odot}}$]       & $5.9^{+1.7}_{-1.2} \times 10^{13}$ \\
M${\rm_{tot}( < 15.4\arcmin = 1}$ Mpc  ) [${\rm M_{\odot}}$]       & $5.9^{+0.8}_{-0.8} \times 10^{14}$ \\
M${\rm_{tot}( < 25.9\arcmin = r_{500}}$) [${\rm M_{\odot}}$]       & $7.9^{+2.5}_{-1.4} \times 10^{14}$  \\
M${\rm_{tot}( < 35\arcmin}$            ) [${\rm M_{\odot}}$]       & $8.8^{+3.4}_{-1.5} \times 10^{14}$  \\
M${\rm_{tot}( < 38.4\arcmin = r_{178}}$) [${\rm M_{\odot}}$]       & $9.1^{+3.6}_{-1.5} \times 10^{14}$   \\
                                                  &   \\
f${\rm_{gas}( < 25.9\arcmin = r_{500}) \times h_{50}^{3/2}}$ & $0.19^{+0.03}_{-0.06}$  \\
f${\rm_{gas}( < 38.4\arcmin = r_{178}) \times h_{50}^{3/2}}$ & $0.26^{+0.05}_{-0.10}$  \\
\noalign{\smallskip\hrule\smallskip}
\end{tabular}
\end{flushleft}
\label{mass_tab}
\end{table}

\section{Conclusions and discussion}
\label{codi}
The heavy element abundace in the central 2\arcmin \ radial bin is significantly above the cluster average. Our values are consistent with the values 
obtained from ASCA data of A3571 (Dupke \& White III 2000a) who also found similar enhancements in clusters A2199 and A496 (Dupke \& White III 2000b).
Also BeppoSAX analyses on several clusters (Irwin \& Bregman 2000; Molendi \& De Grandi 1999; De Grandi \& Molendi 1999b; Molendi et al. 2000)
have yielded central metal enhancements. These observations are consistent with the hypothesis of the metal enrichment of the intracluster gas by supernovae 
(e.g. Arimoto \& Yoshii 1987; Matteucci \& Gibson 1995).
The central BeppoSAX absorption is significantly above the galactic value. Qualitatively similar behavior has been found in other cooling flow clusters  
(e.g. Allen \& Fabian 1997; Molendi et al. 1998; Molendi \& De Grandi, 1999; De Grandi \& Molendi 1999b). This suggests that the excess absorption is due
to the cooling material accumulated in the central region of the cluster. 

The significantly decreasing temperature profile of A3571 derived here with BeppoSAX data agrees with the ASCA temperature profile of Markevitch 
et al. (1998). Similarly, the BeppoSAX temperature profiles for clusters A2029 (Molendi \& De Grandi, 1999), A2256 (Molendi et al. 2000), and A3266 
(De Grandi \& Molendi 1999) show a temperature decline in good agreement with the respective ASCA results (Sarazin et al., 1999; Markevitch \& Vikhlinin, 
1997; Markevitch et al., 1998), although the BeppoSAX profile for A2319 (Molendi et al. 1999) does not indicate a decline reported by Markevitch (1996) from 
ASCA data, even though both give consistent temperatures in the entire overlapping radial range. 
We used both BeppoSAX and ASCA temperature profiles to 
constrain the total mass of A3571, assuming hydrostatic equilibrium. The BeppoSAX profile, even though it is formally consistent with the ASCA one, is 
somewhat steeper and results in problems with convective stability. The best fit cusp model (and the NFW universal profile) for the dark matter density are
too steep compared to the gas density profile for the system to be convectively stable, but they are acceptable within the uncertainties of the data. The 
almost constant temperature up to r = 10\arcmin\ and steep decrease at larger radii, as exhibited with both BeppoSAX and ASCA analysis cannot be acceptably modeled with
polytropic models. The data are well modeled using the ``core'' model for the dark matter density, which is analogous to the $\beta$ model for the gas 
density. The best-fit core model is at the border of being convectively unstable, but can still be accepted. 
The measured masses within r$_{500}$ and r$_{178}$ are factors of 1.1 and 1.4 smaller than the values obtained with the isothermal analysis of A3571 data, respectively, where the difference is consistent with the uncertainties. Within r$_{500}$ the measured mass is significantly smaller, by a
factor of 1.6, than that predicted by the simulations of Evrard et al. (1996).

The dark matter density profile in A3571 decreases as r$^{-6}$, much faster than the gas density. Hence the gas mass fraction increases with radius, with 
f${\rm _{gas}(r_{178}) = 0.26^{+0.05}_{-0.10} \times h_{50}^{3/2}}$. Assuming that this is the lower limit for the primordial baryonic fraction, we obtain 
${\rm \Omega_{m} < 0.4}$ at 90\% confidence.  

\begin{acknowledgements}
The \sax\ satellite is a joint Italian-Dutch programme. 
We thank the staffs of the \sax\ Science Data and
Operations Control Centers for help with these observations. 
J. Nevalainen acknowledges an ESA Research Fellowship.
\end{acknowledgements}

\begin{figure*}
\centerline{\psfig{figure=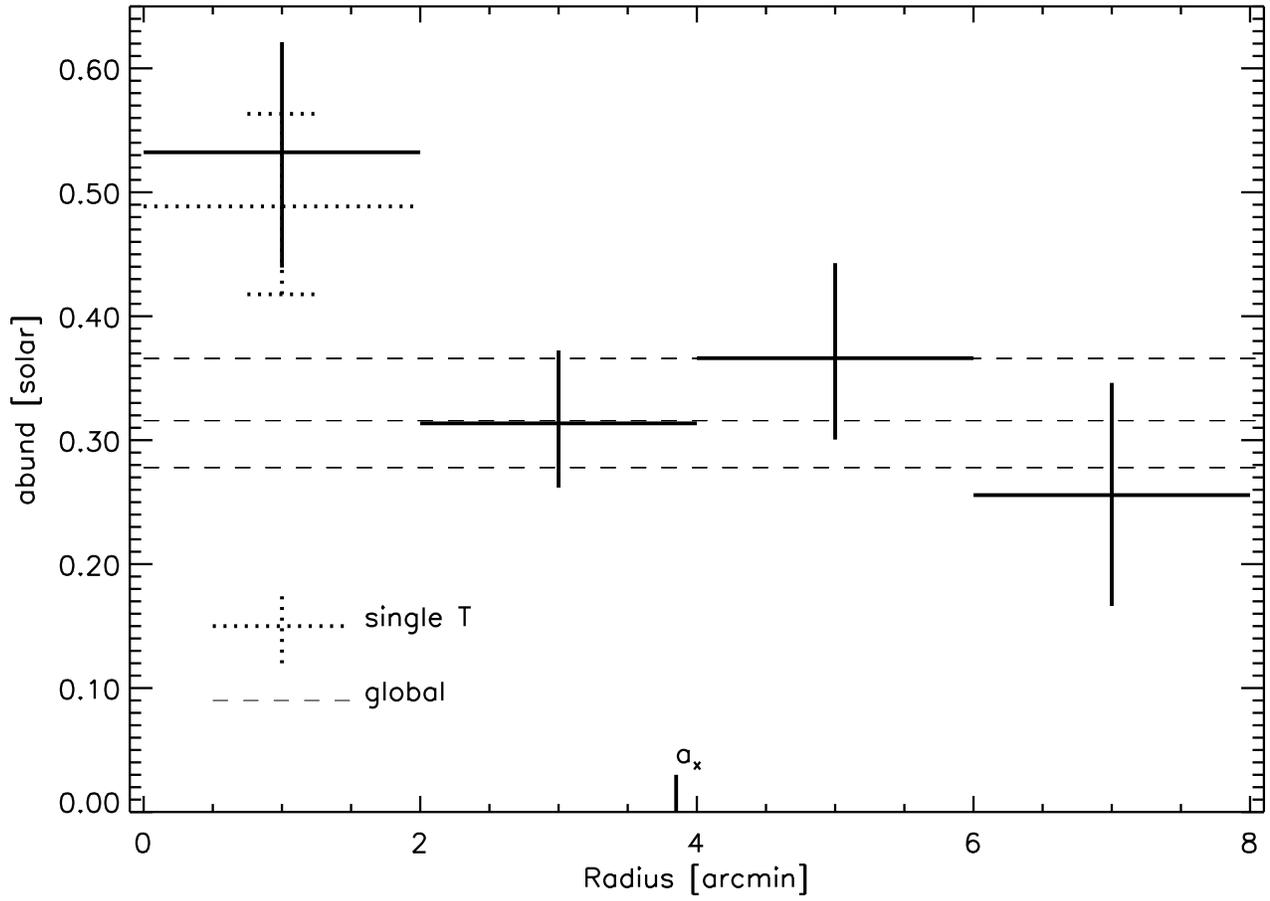,width=18.0cm,angle=0}}
\caption[]{The best-fit values and 90\% confidence uncertainties of the metal abundance corresponding to temperature values in Fig.~\ref{Tplot_fig}. The 
dashed lines show the global values obtained from the fit to the 2\arcmin--8\arcmin\ spectra}
\label{abnhplot_fig}
\end{figure*}

\begin{figure*}
\centerline{\psfig{figure=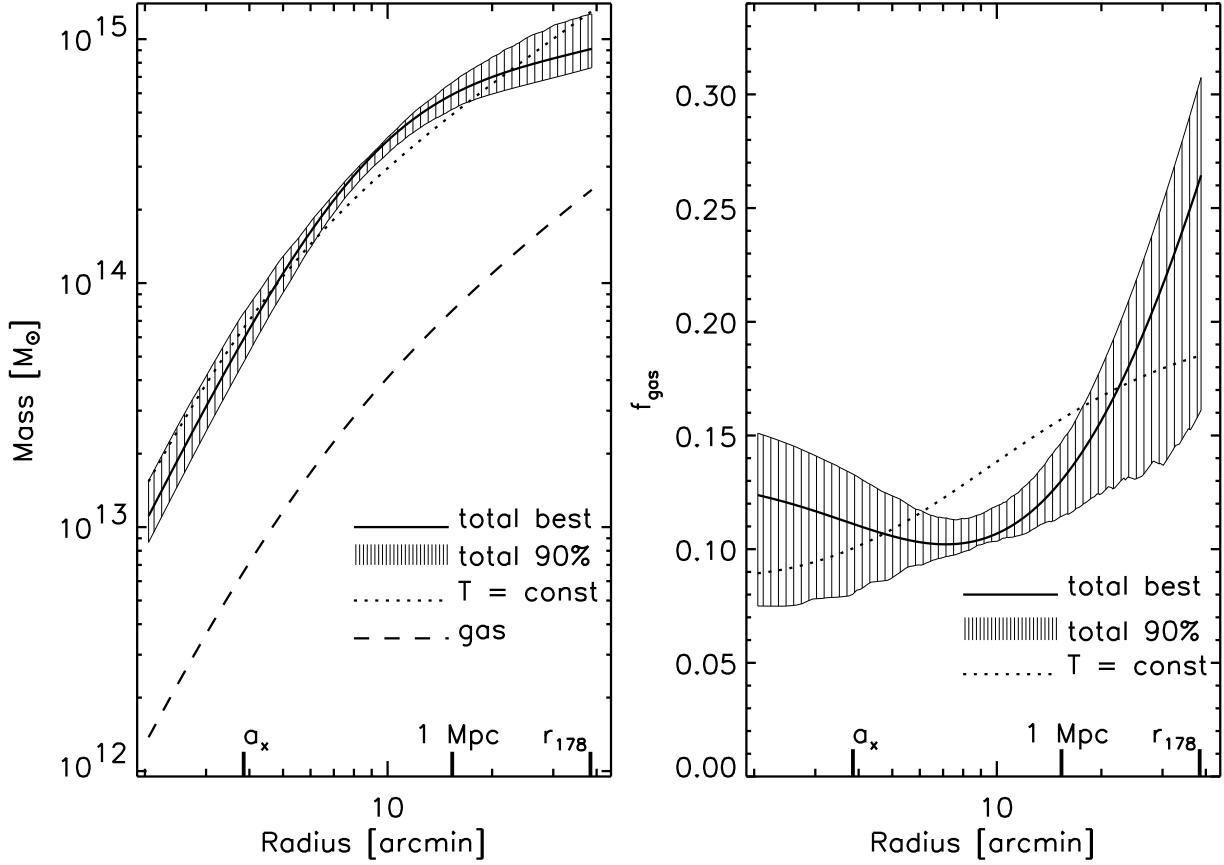,width=18.0cm,angle=0}}
\caption[]{The left panel shows the total mass profile (solid line) with 90\% confidence uncertainties (shaded area) obtained with the core model. The dotted
line shows the total mass obtained assuming a constant temperature $T = 6.9$ keV along the cluster. The dashed line shows the gas mass profile. The right 
panel shows the corresponding gas mass fraction profiles. Masses are evaluated using ${\rm H_{0}} = 50$ km s$^{-1}$ Mpc$^{-1}$ (total mass scales as 
H$^{-1}$ and gas mass as H$^{-5/2}$)}
\label{massplot_fig}
\end{figure*}

\end{document}